\documentstyle[12pt]{article}

\def\hybrid{\topmargin -20pt    \oddsidemargin 0pt
        \headheight 0pt \headsep 0pt
        \textwidth 6.25in       
        \textheight 9.5in       
        \marginparwidth .875in
        \parskip 5pt plus 1pt   \jot = 1.5ex}
\def\cQ{{\cal Q}}
\def\cG{{\cal G}}
\def\cL{{\cal L}}
\def\cH{{\cal H}}
\def\ket#1{|{#1}\rangle}
\def\noi{\noindent}
\def\half{{1\over2}}
\def\baselinestretch{1.2}

\catcode`\@=11

\def\marginnote#1{}
\def\draftlabel#1{{\@bsphack\if@filesw {\let\thepage\relax
   \xdef\@gtempa{\write\@auxout{\string
      \newlabel{#1}{{\@currentlabel}{\thepage}}}}}\@gtempa
   \if@nobreak \ifvmode\nobreak\fi\fi\fi\@esphack}
        \gdef\@eqnlabel{#1}}
\def\@eqnlabel{}
\def\@vacuum{}
\def\draftmarginnote#1{\marginpar{\raggedright\scriptsize\tt#1}}

\def\draft{\oddsidemargin -.2truein
        \def\@oddfoot{\sl preliminary draft \hfil
        \rm\thepage\hfil\sl\today\quad\militarytime}
        \let\@evenfoot\@oddfoot \overfullrule 3pt
        \let\label=\draftlabel
        \let\marginnote=\draftmarginnote
   \def\@eqnnum{(\theequation)\rlap{\kern\marginparsep\tt\@eqnlabel}%
\global\let\@eqnlabel\@vacuum}  }

\def\preprint{\twocolumn\sloppy\flushbottom\parindent 2em
        \leftmargini 2em\leftmarginv .5em\leftmarginvi .5em
        \oddsidemargin -.5in    \evensidemargin -.5in
        \columnsep .4in \footheight 0pt
        \textwidth 10.in        \topmargin  -.4in
        \headheight 12pt \topskip .4in
        \textheight 6.9in \footskip 0pt
        \def\@oddhead{\thepage\hfil\addtocounter{page}{1}\thepage}
        \let\@evenhead\@oddhead \def\@oddfoot{} \def\@evenfoot{} }

\def\numberbysection{\@addtoreset{equation}{section}
        \def\theequation{\thesection.\arabic{equation}}}

\def\underline#1{\relax\ifmmode\@@underline#1\else
        $\@@underline{\hbox{#1}}$\relax\fi}

\def\titlepage{\@restonecolfalse\if@twocolumn\@restonecoltrue
\onecolumn
     \else \newpage \fi \thispagestyle{empty}\c@page\z@
        \def\thefootnote{\fnsymbol{footnote}} }

\def\endtitlepage{\if@restonecol\twocolumn \else \newpage \fi
        \def\thefootnote{\arabic{footnote}}
        \setcounter{footnote}{0}}  

\catcode`@=12
\relax

\def\figcap{\section*{Figure Captions\markboth
        {FIGURECAPTIONS}{FIGURECAPTIONS}}\list
        {Figure \arabic{enumi}:\hfill}{\settowidth\labelwidth{Figure
999:}
        \leftmargin\labelwidth
        \advance\leftmargin\labelsep\usecounter{enumi}}}
 \relax
\def\tablecap{\section*{Table Captions\markboth
        {TABLECAPTIONS}{TABLECAPTIONS}}\list
        {Table \arabic{enumi}:\hfill}{\settowidth\labelwidth{Table
999:}
        \leftmargin\labelwidth
        \advance\leftmargin\labelsep\usecounter{enumi}}}
 \relax
\def\reflist{\section*{References\markboth
        {REFLIST}{REFLIST}}\list
        {[\arabic{enumi}]\hfill}{\settowidth\labelwidth{[999]}
        \leftmargin\labelwidth
        \advance\leftmargin\labelsep\usecounter{enumi}}}
 \relax
\makeatletter
\newcounter{pubctr}
\def\publist{\@ifnextchar[{\@publist}{\@@publist}}
\def\@publist[#1]{\list
        {[\arabic{pubctr}]\hfill}{\settowidth\labelwidth{[999]}
        \leftmargin\labelwidth
        \advance\leftmargin\labelsep
        \@nmbrlisttrue\def\@listctr{pubctr}
        \setcounter{pubctr}{#1}\addtocounter{pubctr}{-1}}}
\def\@@publist{\list
        {[\arabic{pubctr}]\hfill}{\settowidth\labelwidth{[999]}
        \leftmargin\labelwidth
        \advance\leftmargin\labelsep
        \@nmbrlisttrue\def\@listctr{pubctr}}}
 \relax
\makeatother
\newskip\humongous \humongous=0pt plus 1000pt minus 1000pt

\newif\ifdtup

\font\Scbig=cmss10 scaled\magstep1
\font\Scscr=cmss8 scaled\magstep1
\font\Scscrscr=cmss8
\newfam\Scfam
\textfont\Scfam=\Scbig
\scriptfont\Scfam=\Scscr
\scriptscriptfont\Scfam=\Scscrscr
\def\Sc{\fam\Scfam}

\relax
\hybrid
\def\lvm{\leavevmode\hbox to\parindent{\hfill}}

\def\thefootnote{\fnsymbol{footnote}}
\def\BE{\begin{equation}}
\def\EE{\end{equation}}
\def\BA{\begin{eqnarray}}
\def\EA{\end{eqnarray}}

\def\th{\theta}

\def\tt{\bar\tau}

\def\lvm{\leavevmode\hbox to\parindent{\hfill}}
\def\bar{\overline}
\def\req#1{(\ref{#1})}

\def\BE{\begin{equation}}
\def\EE{\end{equation} \vskip 0.30\baselineskip}
\def\BA{\begin{array}}
\def\EA{\end{array}}

\def\noi{\noindent}

\def\frac#1#2{{\textstyle{{#1}\over{#2}}}}
\def\half{{1\over2}}

\def\ket#1{|{#1}\rangle}

\def\cA{{\cal A}}
\def\cG{{\cal G}}
\def\cH{{\cal H}}

\def\cL{{\cal L}}

\def\cQ{{\cal Q}}

\def\cU{{\cal U}}

\def\ctop{{\Sc c}}

\def\ie{{\it i.e.}}

\def\Qz{\cQ_0}
\def\Gz{\cG_0}
\def\Qn{$\Qz$}
\def\Gn{$\Gz$}
\def\kc{{\ket{\chi}}}

\newif\ifold \oldtrue \def\new{\oldfalse}
\let\ssection=\section
\def\section{\setcounter{equation}{0}\ssection}

\begin{document}
\renewcommand{\theequation}{\thesection.\arabic{equation}}
\newcommand{\beq}{\begin{equation}}
\newcommand{\eeq}[1]{\label{#1}\end{equation}}
\newcommand{\ber}{\begin{eqnarray}}
\newcommand{\eer}[1]{\label{#1}\end{eqnarray}}
\begin{titlepage}
\begin{center}

\hfill IMAFF-95/37\\
\hfill NIKHEF-95-063\\
\hfill hep-th/9511208\\
\vskip .6in

{\large \bf  The Other Spectral Flow}
\vskip .8in

{\bf Beatriz Gato-Rivera}$^{a,b}$ {\bf and Jose Ignacio Rosado}$^a$\\
\vskip .3in

${\ }^a$
{\em Instituto de Matem\'aticas y
 F\'\i sica Fundamental, CSIC,\\ Serrano 123,
Madrid 28006, Spain} \footnote{e-mail addresses:
bgato, jirs @pinar1.csic.es}\\
\vskip .25in

${\ }^b$
 {\em NIKHEF-H Kruislaan 409, NL-1098 SJ Amsterdam, The Netherlands}\\

\vskip .5in

\end{center}

\vskip .6in

\begin{center} {\bf ABSTRACT } \end{center}
\begin{quotation}
Recently we showed that the spectral flow acting on the N=2 twisted
topological theories gives rise to a topological algebra automorphism.
Here we point out that the untwisting of that automorphism leads to
a spectral flow on the untwisted
N=2 superconformal algebra which is different from the usual one.
 This ``other" spectral flow does not
interpolate between the chiral ring and the antichiral ring.
 In particular it maps the chiral ring
into the chiral ring and the antichiral ring into the antichiral ring.
We discuss the similarities and differences between both spectral
flows. We also analyze their action on null states.

\end{quotation}
\vskip 1.5cm

November 1995\\
\end{titlepage}
\vfill
\eject
\def\baselinestretch{1.2}
\baselineskip 17 pt
\section{Introduction}\lvm

In a recent paper \cite{BJI3} we analyzed the action of the
 spectral flow on the topological theories obtained by twisting
a N=2 superconformal theory. We found that the spectral flow
interpolates between the two twisted theories corresponding to
a given superconformal one in the following way

\BE\new\BA{rclcrcl}
\cU_1 \, \cL^{(2)}_m \, \cU_1^{-1}&=& \cL_m^{(1)} - m\cH_m^{(1)}\,,\\
\cU_1 \, \cH^{(2)}_m \, \cU_1^{-1}&=&-\cH_m^{(1)}
 -{\ctop\over3} \delta_{m,0}\,,\\
\cU_1 \, \cQ^{(2)}_m \, \cU_1^{-1}&=&\cG_m^{(1)}\,,\\
\cU_1 \, \cG^{(2)}_m \, \cU_1^{-1}&=&\cQ_m^{(1)}\,,\
\label{spfla}\EA\EE

\noi
where (1) and (2) denote the topological generators.
They are obtained
through the twistings (1) and (2) of the superconformal generators,
namely

\BE\new\BA{rclcrcl}
\cL^{(1)}_m&=&\multicolumn{5}{l}{L_m+\half(m+1)H_m\,,}\\
\cH^{(1)}_m&=&H_m\,,&{}&{}&{}&{}\\
\cG^{(1)}_m&=&G_{m+\half}^+\,,&\qquad &\cQ_m^{(1)}&=&G^-_{m-\half}
\,,\label{twa}\EA\EE

\noi
and

\BE\new\BA{rclcrcl}
\cL^{(2)}_m&=&\multicolumn{5}{l}{L_m-\half(m+1)H_m\,,}\\
\cH^{(2)}_m&=&-H_m\,,&{}&{}&{}&{}\\
\cG^{(2)}_m&=&G_{m+\half}^-\,,&\qquad &
\cQ_m^{(2)}&=&G^+_{m-\half}\,,\label{twb}\EA\EE

\noi
and $\cU_1$ is the spectral flow operator $\cU_{\th}$ for
$\th=1$ (one gets the same transformation \req{spfla} exchanging
$(1) \leftrightarrow (2)$ and $\cU_1 \leftrightarrow \cU_{-1}$).

Observe that the topological chiral primaries, \ie\ those annihilated
by \Qn\ and \Gn, are the superconformal chiral primaries
belonging to the chiral ring and
the  antichiral ring, for the twists (2) and
(1) respectively.

In ref. \cite{BJI3} we argued that, as long as we stay at the
topological algebra level without going into specific
realizations of the twisted topological theories, we can
regard \req{spfla}, without labels (1) and (2), as a topological algebra
automorphism. The reason is that both sets of generators satisfy
exactly the same algebra, so that at the topological algebra
level they are indistinguishable from each other. This
topological algebra automorphism was expressed introducing a
new operator $\cA$ in the form

\BE\new\BA{rclcrcl}
\cA \, \cL_m \, \cA&=& \cL_m - m\cH_m\,,\\
\cA \, \cH_m \, \cA&=&-\cH_m - {\ctop\over3} \delta_{m,0}\,,\\
\cA \, \cQ_m \, \cA&=&\cG_m\,,\\
\cA \, \cG_m \, \cA&=&\cQ_m\,,\
\label{autom} \EA\EE

\noi
with $\cA^{-1} = \cA$.

In this note we intend to bring to the reader's attention
the fact that the untwisting of the topological algebra
automorphism \req{autom} does not give back the ``usual"
spectral flow on the N=2 superconformal generators, but
a different transformation: the ``other" spectral flow.
In what follows we will discuss the similarities and differences
between the two versions of the spectral flow.

\section{The Usual Spectral Flow}\lvm

Let us start with the usual spectral flow \cite{SS}, \cite{LVW}.
It is given by the one-parameter family of transformations

\BE\new\BA{rclcrcl}
\cU_\th \, L_m \, \cU_\th^{-1}&=& L_m
 +\th H_m + {\ctop\over 6} \th^2 \delta_{m,0}\,,\\
\cU_\th \, H_m \, \cU_\th^{-1}&=&H_m + {\ctop\over3} \th \delta_{m,0}\,,\\
\cU_\th \, G^+_r \, \cU_\th^{-1}&=&G_{r+\th}^+\,,\\
\cU_\th \, G^-_r \, \cU_\th^{-1}&=&G_{r-\th}^-\,.\
\label{spfl} \EA\EE

\noi
on the generators of the N=2 superconformal algebra \cite{Adem},
 \cite{DiVPZ}, satisfying $\cU^{-1}_\th = \cU_{(-\th)}$.
These transformations give rise to isomorphic algebras.

In order to simplify the analysis that follows, it will be
very convenient to unify the notation for the $U(1)$
charge of the states of the Neveu-Schwarz (aperiodic)
algebra and the states of the Ramond (periodic) algebra.
Namely, the $U(1)$ charge of the Ramond states will be
denoted by $h$, instead of $h\pm \half$. In addition, the
relative charge of a secondary state $q$ will be defined as
the difference between the $U(1)$ charge of the state and the
$U(1)$ charge of the primary on which it is built.
Therefore, the relative charges of the Ramond states are
defined to be integer (in contrast with the usual definition).

Let us denote by $(\Delta, h)$ the $(L_0, H_0)$
eigenvalues of any given state $\kc$, then it is easy to see
that the eigenvalues of the transformed state
$\cU_{\th} \kc$ are
 $(\Delta-\th h +{c\over6} \th^2, h- {c\over3} \th)$.
If the state $\kc$ is a level-$l$ secondary state with relative
 charge $q$ and eigenvalues $(\Delta+l, h+q)$ (where now
$(\Delta, h)$ denote the eigenvalues of the primary on
which the secondary is built), then one gets straightforwardly
that the level of the transformed state $\cU_{\th} \kc$ changes
to $l-\th q$, while the relative charge remains equal.

In the discussion that follows we will assume the same behaviour
for left-movers and right-movers, so that no distinction
between them will be made.

 For half-integer values
of $\th$ the spectral flow \req{spfl} interpolates between
the NS algebra and the R
algebra. In particular, for $\th = \half$ the
primaries of the NS algebra become primaries of the R
 algebra, with chirality $(-)$(\ie\ annihilated by $G^-_0$),
and the chiral ring (the primaries of the NS algebra
annihilated by $G^+_{-1/2}$) is mapped into the set of
Ramond ground states (annihilated by both $G^+_0$ and
$G^-_0$). Similarly, for $\th = -\half$ the
primaries of the NS algebra are transformed into primaries
of the R algebra with chirality $(+)$ and the antichiral ring
(primaries annihilated by $G^-_{-1/2}$) is mapped into the
Ramond ground states.

For integer values of $\th$ the N=2 superconformal
algebras map back to themselves, although the primary fields
change, \ie\ the fields that were primary with respect to
the initial algebra are in general not longer primary
with respect to the final algebra. Only in particular cases
the spectral flow maps the initial primary states into
primary states as well. For example, in the case of
 the NS algebra only primary states that are chiral
(or antichiral) can be mapped into primary states of the
NS algebra. In particular,
 for  $\th =1$ the
chiral ring is mapped into the antichiral ring (while
the antichiral ring is mapped to a set of non-primary
fields), and for $\th = -1$ the antichiral
ring is mapped into the chiral ring (while the chiral
ring is mapped into non-primary fields).

In the case of the R algebra, the spectral flow
with $\th = 1$ transforms primary states with chirality $(+)$
into primary states with chirality $(-)$, while primary states
with chirality $(-)$ become non-primaries and annihilated by
$G^-_{-1}$. Therefore, the Ramond ground states are transformed
under $\cU_1$ into chirality $(-)$ primaries
with the additional
condition of being annihilated by $G^-_{-1}$. Similarly,
for $\th = -1$ the primary states with chirality $(-)$ are
transformed into primary states with chirality $(+)$ and
the ground states are mapped into chirality $(+)$ primaries
with the additional condition of being annihilated by $G^+_{-1}$.

\section{The Other Spectral Flow}\lvm

The untwisting of the topological algebra automorphism
\req{autom}, under the twist (1), eq. \req{twa}, gives the transformation

\BE\new\BA{rclcrcl}
\cA^{(1)} \, L_m \, \cA^{(1)}&=& L_m
 + H_m + {\ctop\over 6} \delta_{m,0}\,,\\
\cA^{(1)} \, H_m \, \cA^{(1)}&=&- H_m - {\ctop\over3} \delta_{m,0}\,,\\
\cA^{(1)} \, G^+_r \, \cA^{(1)}&=&G_{r-1}^-\,,\\
\cA^{(1)} \, G^-_r \, \cA^{(1)}&=&G_{r+1}^+\,,\\
\label{ospa} \EA\EE

\noi
while under the twist (2), eq. \req{twb},  it results in

\BE\new\BA{rclcrcl}
\cA^{(2)} \, L_m \, \cA^{(2)}&=& L_m
 - H_m + {\ctop\over 6} \delta_{m,0}\,,\\
\cA^{(2)} \, H_m \, \cA^{(2)}&=&- H_m + {\ctop\over3} \delta_{m,0}\,,\\
\cA^{(2)} \, G^+_r \, \cA^{(2)}&=&G_{r+1}^-\,,\\
\cA^{(2)} \, G^-_r \, \cA^{(2)}&=&G_{r-1}^+\,,\
\label{ospb} \EA\EE

\noi
where $\cA^{(1)}$ and $\cA^{(2)}$ denote the ``untwisted"
automorphisms under the twists (1) and (2) respectively.

One can therefore regard the transformations \req{ospa} and
\req{ospb} as the particular cases, for $\th = 1$ and $\th = -1$
respectively, of a different spectral flow given by

\BE\new\BA{rclcrcl}
\cA_\th \, L_m \, \cA_\th&=& L_m
 +\th H_m + {\ctop\over 6} \th^2 \delta_{m,0}\,,\\
\cA_\th \, H_m \, \cA_\th&=&- H_m - {\ctop\over3} \th \delta_{m,0}\,,\\
\cA_\th \, G^+_r \, \cA_\th&=&G_{r-\th}^-\,,\\
\cA_\th \, G^-_r \, \cA_\th&=&G_{r+\th}^+\,.\
\label{ospfl} \EA\EE

Observe that $\cU_{\th}^{-1} = \cU_{(-\th)}$ while
$\cA_{\th}^{-1} = \cA_{\th}$, \ie\ the inverse of the usual
spectral flow with parameter $\th$ is the one given by $(-\th)$
while for the other spectral flow the transformation and
its inverse coincide. The $(L_0, H_0)$ eigenvalues of the
transformed states
$\cA_{\th} \kc$ are now
 $(\Delta+\th h +{c\over6} \th^2, - h - {c\over3} \th)$
(that is, they differ from the previous case by the
 interchange $h \rightarrow -h$). From this one easily
deduces that, under the spectral flow \req{ospfl}, the level $l$
of any descendant will change to $l + \th q$ while
the relative charge $q$ reverse its sign.

Observe also that, due to the untwisting procedure, the natural
Hilbert space of the transformation given by \req{ospa} is
the antichiral ring, while the natural Hilbert space of
the transformation given by \req{ospb} is the chiral ring.
As a matter of fact, \req{ospa} maps the antichiral ring
into the antichiral ring (transforming the chiral ring
into a set of non-primary fields),
 while \req{ospb} maps the chiral ring into the chiral ring
(and the antichiral ring
into non-primary fields). Not only
that, but for no values of $\th$ does the
spectral flow \req{ospfl} interpolate between the chiral
ring and the antichiral ring. In addition, as happened with
the usual spectral flow, only primaries of the NS algebra
 that are chiral or
 antichiral can be mapped into primaries of the NS algebra.

For half-integer values of $\th$ the spectral flow \req{ospfl}
 also interpolates between the NS algebra and the R algebra.
For $\th = \half$ the
primaries of the NS algebra become primaries of the R
 algebra with chirality $(-)$, as before, but now it is
the antichiral ring which
is mapped into the
Ramond ground states.
Similarly, for $\th = -\half$ the
primaries of the NS algebra are transformed into primaries
of the R algebra with chirality $(+)$, while now the chiral ring
is mapped into the
Ramond ground states.

In the case of the R algebra, the spectral flow \req{ospfl}
with $\th = 1$ transforms primary states with chirality $(-)$
back into chirality $(-)$ primaries; that is, $\cA_1$ does
not reverse the chirality as $\cU_1$ does. Primary states with
chirality $(+)$, in turn, are transformed
into non-primary states annihilated by $G^-_{-1}$.
Therefore, the Ramond ground states are transformed
under $\cA_1$ into chirality $(-)$ primaries
with the additional
condition of being annihilated by $G^-_{-1}$.
Similarly,
for $\th = -1$ the primary states with chirality $(+)$ are
transformed back into primary states with chirality $(+)$ and
the ground states are mapped into chirality $(+)$ primaries
with the additional condition of being annihilated by $G^+_{-1}$.

\section{Spectral Flows on Null States}\lvm

Now let us discuss the behaviour of the spectral flows
\req{spfl} and \req{ospfl} when they act on
 null states; that is,
descendant states that also satisfy the highest weight
conditions

$$   L_{n>0} \kc =  H_{n>0} \kc =
 G^+_{n>0} \kc =  G^-_{n>0} \kc  = 0 $$

\noi
which define primary states. Since null states are also primaries,
we can apply the analysis above regarding the action of the spectral
flows on primary states. However, we have to take into account
that null states of the NS algebra cannot
be chiral or antichiral. In particular, this applies to null
states built on the chiral ring and
the antichiral ring, \ie\ they cannot
be chiral or antichiral themselves (a detailed analysis of this
fact will be given in \cite{BJI5}). Also we have to take into
 account that the possible values of the relative charge $q$
are very limited for null states: zero and $\pm 1$ \cite{BFK}
(in the usual notation $\pm \half$ and
 $\pm {3\over2}$ for the null vectors of the Ramond algebra).

\vskip .2in

The main features about the action of the spectral flows
\req{spfl} and \req{ospfl} on null states, as deduced from
the above discussions, are therefore the following, where
$\kc_l^{(q)}$ denotes a level-$l$ null state with
relative charge $q$.

*) The spectral flows \req{spfl} and \req{ospfl}
 map null states $\kc_l^{(q)}$
 of the NS algebra into
null states of the R algebra (and viceversa),
 for $\th = \pm \half$. The $(L_0, H_0)$ eigenvalues
of the primaries on which the R null states are built
are given by $(\Delta \pm \half h + {c \over 24}, h \pm {c\over6})$,
using the usual spectral flow, and
$(\Delta \pm \half h + {c \over 24},-( h \pm {c\over6}))$, using
the other spectral flow, where $(\Delta, h)$ are the eigenvalues
of the primaries on which the NS null states are built.
The levels of the R null states are given by
$l \pm \half q$ , the sign depending on the sign of $\th$ and the
specific spectral flow, and the relative charges are
given by $q$ (usual spectral flow) and $-q$ (other spectral
flow). The chiralities of the R null states are $(-)$ for
$\th = \half$ and $(+)$ for $\th = - \half$.

*) For no value of $\th$ null states of the NS
 algebra are mapped into null states
of the NS algebra. This applies in particular to null states
built on the chiral ring and the antichiral ring.

*) Null states of the R algebra are mapped into null states of
the R algebra, for $\th= \pm 1$.
The $(L_0, H_0)$ eigenvalues of the primaries on which the
null states are built change from $(\Delta, h)$ to
$(\Delta \pm  h + {c \over 6}, h \pm {c\over3})$,
using the usual spectral flow, and
to $(\Delta \pm  h + {c \over 6},  -(h \pm {c\over3}))$,
using the other spectral flow.
 The level of the null states
gets modified to $l \pm q$, the sign depending
 on the sign of $\th$ and
the spectral flow at hand,
while only the other spectral flow
modifies the relative charge, reversing its sign.
In addition, the usual spectral flow reverses the chirality,
while the other spectral flow does not.

\section{Final Remarks}\lvm

We have shown that the ``usual" spectral flow $\cU_\th$
 and the ``other"
spectral flow $\cA_\th$, although similar in
 many of their properties, differ notably
in some respects. Their behaviour with
respect to the chiral ring and the antichiral ring is
drastically different, and complementary. The usual
spectral flow interpolates between the chiral ring and the
 antichiral ring, but does not connect the chiral ring
and the antichiral ring back to themselves.
 The other spectral flow
does exactly the opposite, mapping the chiral ring and the
antichiral ring back to themselves, while being unable to
connect the two rings to each other. Furthermore their
inverses are different  : $\cU^{-1}_\th = \cU_{(-\th)}$
and $\cA^{-1}_\th = \cA_{\th}$.

 We have already found a practical application
of this result. As we discussed above, the spectral
flows themselves do not map null states of the NS algebra
into null states of the NS algebra. However,
using the two spectral flows we have
constructed operators that map null states of the NS
algebra  into null states of the NS algebra
 \cite{BJI5}, \cite{BJI6}. In
particular, using the usual spectral flow, the resulting
operators transform null states built on the chiral ring
into null states built on the antichiral ring (and
viceversa).
 On the other hand, using the other spectral
flow, the corresponding operators map null states built on
the chiral ring into null states built on the chiral ring,
and null states built on the antichiral ring into null
states built on the antichiral ring.

We think that other applications will be found for the ``other"
spectral flow described here.

\end{document}